\documentclass[useAMS,usenatbib]{mn2e}

\usepackage{graphicx}
\newcommand{\bm}{\bmath}
 
\title[Yukawa effects on the clock onboard a drag-free satellite]{Yukawa effects on the clock onboard a drag-free satellite}
\author[X.-M. Deng \& Y. Xie]{Xue-Mei Deng$^{1,3}$ and Yi Xie$^{2,3}$\thanks{To whom correspondence should be addressed. E-mail: yixie@nju.edu.cn}\\
$^{1}$Purple Mountain Observatory, Chinese Academy of Sciences, Nanjing 210008, China\\
$^{2}$Department of Astronomy, Nanjing University, Nanjing 210093, China\\
$^{3}$Key Laboratory of Modern Astronomy and Astrophysics, Nanjing University, Ministry of Education, Nanjing 210093, China}
\begin{document}

\date{Accepted . Received ; in original form }

\pagerange{\pageref{firstpage}--\pageref{lastpage}} \pubyear{2002}

\maketitle

\label{firstpage}

\begin{abstract}
The Yukawa correction to the Newtonian gravitational force is accepted as a parameterization of deviations from the inverse-square law of gravity which might be caused by new physics beyond the standard model of particles and the general theory of relativity. We investigate these effects on the clock onboard a drag-free satellite: dynamics of the satellite and influence on the time transfer link.  We find the Yukawa signal in the time transfer is much harder to detect with current state of clocks than those effects on the dynamics, especially the secular change of periastron, by laser ranging in the case of an artificial Earth satellite carrying a frequency standard with an orbit of $a=10^7$\,m and $e=0.01$.
\end{abstract}


\section{Introduction}           

It is gravitation that is the first of the four known fundamental forces to be understood quantitatively and empirically. However, unlike the strong, weak and electromagnetic interactions, it is also gravitation that still can not be explained as a result of the quantum exchange of virtual bosons. Theoretically, despite some candidate theories proposed in recent years, unification of gravitation and the rest part of physics remains as a grand challenge. Experimentally and observationally, searching for possible deviations from the inverse-square law of gravity [see \cite{Adelberger2003} for a review], which are predicted by some possible schemes of quantum gravity, might shed light on new physics. One of the parameterizations of the deviations is the Newtonian gravitational potential with an additional Yukawa correction \citep{Fischbach1986,Fischbach1992} that is
\begin{equation}
  \label{}
  V = V_{\mathrm{N}}(r)+V_{\mathrm{YK}}(r),
\end{equation}
where
\begin{eqnarray}
  \label{}
  V_{\mathrm{N}}(r) & = & \frac{Gm_1m_2}{r},\\
  \label{VYK}
  V_{\mathrm{YK}}(r) & = & \frac{Gm_1m_2}{r}\alpha \exp\bigg(-\frac{r}{\lambda}\bigg).
\end{eqnarray}
Here $\alpha$ is a dimensionless strength parameter and $\lambda$ is a length scale [see \cite{Fischbach1999} for a review of constraints on $\alpha$ and $\lambda$] .

Inspired by the idea of orbital tests of relativistic gravity using Earth satellites \citep{Damour1994}, we would like to investigate the Yukawa effects on the clock onboard a satellite and their observability. In this methodology \citep{Damour1994}, the satellite will use drag-free technology \citep{Pugh1959,Lange1964}, making it be able to follow a free fall trajectory, and its three-dimensional motion is tracked at the centimeter or even less level either by laser ranging or the global positioning system. The advantage of using an space-borne clock is that Yukawa effects, in principle, will appear not only in the dynamics of the satellite but also in the time transfer signals.

In Section \ref{YukawaSat}, treating the Yukawa correction as small disturbance, we will follow the standard procedure of perturbation in the celestial mechanics and obtain the Lagrange planetary equations of the satellite. It is worth mentioning that \cite{Haranas2011} derived the same equations but their results contradict the conservation law of the angular momentum in the central force problem (see Sec. \ref{YukawaSat} for details). We will study the Yukawa effects on the time transfer of the onboard clock in Section \ref{YukawaTT}. In Section \ref{obs}, the observability of these effects is discussed. Conclusions and discussion will be debriefed in Section \ref{condis}.

\section{Yukawa effects on satellite dynamics}

\label{YukawaSat}

There is no doubt that an artificial Earth satellite practically feels a lot of non-gravitational disturbance, such as atmospheric drags and radiation pressures. These effects are usually larger or much larger than the relativistic corrections to the Newtonian gravity and some effects caused by possible deviations from the inverse-square law of gravity. A very promising way to suppress them to a tiny level is the drag-free technology \citep{Pugh1959,Lange1964} that is already employed by the Gravity Probe B (GP-B) \citep{Everitt2011PRL106.221101} and Gravity field and steady-state Ocean Circulation Explorer (GOCE) \citep{Drinkwater2003SSR108.419}. Its basic principle is that a space vehicle offsets the non–gravitational orbital disturbances by staying around an inside test mass that is shielded from the environment. Since the test mass is governed by the gravitational force only, the satellite is able to follow a free fall trajectory in the background gravitational field. In the following parts of this work, we assume that the clock used to detect the Yukawa effects will be carried onboard such a drag-free satellite so that we focus on the pure gravitational interactions. Since the relativistic celestial mechanics of a satellite has been intensively and well studied \citep{Brumberg1989,Soffel1989,Brumberg1991,Damour1994PRD49.618,Kopeikin2011}, we will pay attention to the Yukawa effects only.

With widely used notations in celestial mechanics, if $a$, $e$, $i$, $\Omega$, $\omega$ and $M$ are taken as independent orbital arguments, the evolution of a satellite orbit under a perturbing potential $R$ is \citep{Danby1962}
\begin{eqnarray}
\frac{\mathrm{d}a}{\mathrm{d}t}&=&\frac{2}{na}\frac{\partial R}{\partial M},\\
\frac{\mathrm{d}e}{\mathrm{d}t}&=&\frac{(1-e^{2})}{na^{2}e}\frac{\partial R}{\partial M}-\frac{\sqrt{1-e^{2}}}{na^{2}e}\frac{\partial R}{\partial\omega},\\
\frac{\mathrm{d}i}{\mathrm{d}t}&=&\frac{\cos i}{na^{2}\sqrt{1-e^{2}}\sin i}\frac{\partial R}{\partial\omega}-\frac{1}{na^{2}\sqrt{1-e^{2}}\sin i}\frac{\partial R}{\partial\Omega},\\
\frac{\mathrm{d}\Omega}{\mathrm{d}t}&=&\frac{1}{na^{2}\sqrt{1-e^{2}}\sin i}\frac{\partial R}{\partial i},\\
\frac{\mathrm{d}\omega}{\mathrm{d}t}&=&\frac{\sqrt{1-e^{2}}}{na^{2}e}\frac{\partial R}{\partial e}-\frac{\cos i}{na^{2}\sqrt{1-e^{2}}\sin i}\frac{\partial R}{\partial i},\\
\frac{\mathrm{d}M}{\mathrm{d}t}&=&n-\frac{(1-e^{2})}{na^{2}e}\frac{\partial R}{\partial e}-\frac{2}{na}\frac{\partial R}{\partial a}.
\end{eqnarray}
In the case of Yukawa correction to the Newtonian potential, according to Eq. (\ref{VYK}),
\begin{equation}
  \label{}
  R = \frac{\mu}{r}\alpha \exp\bigg(-\frac{r}{\lambda}\bigg),
\end{equation}
where $\mu\equiv GM$. The above expression shows $R$ depends on $r$ only, i.e. $R=R(r)$, which means the two-body problem under the framework of the Newtonian gravitation with the Yukawa correction is a special case of the central force problem. Evaluating the partial derivatives involving $R$, we have the Lagrange planetary equations for the satellite as
\begin{eqnarray}
\frac{\mathrm{d}a}{\mathrm{d}t}&=&-\frac{2e}{n\sqrt{1-e^{2}}}\frac{\mu}{r^{2}}\alpha\exp\bigg(-\frac{r}{\lambda}\bigg)\bigg(1+\frac{r}{\lambda}\bigg) \sin f,\\
\frac{\mathrm{d}e}{\mathrm{d}t}&=&-\frac{\sqrt{1-e^{2}}}{na}\frac{\mu}{r^{2}}\alpha\exp\bigg(-\frac{r}{\lambda}\bigg)\bigg(1+\frac{r}{\lambda}\bigg) \sin f,\\
\frac{\mathrm{d}i}{\mathrm{d}t}&=&0,\\
\frac{\mathrm{d}\Omega}{\mathrm{d}t}&=&0,\\
\frac{\mathrm{d}\omega}{\mathrm{d}t}&=&\frac{\sqrt{1-e^{2}}}{nae}\frac{\mu}{r^{2}}\alpha\exp\bigg(-\frac{r}{\lambda}\bigg)\bigg(1+\frac{r}{\lambda}\bigg)
\cos f,\\
\frac{\mathrm{d}M}{\mathrm{d}t}&=&n-\frac{(1-e^{2})}{nae}\frac{\mu}{r^{2}}\alpha\exp\bigg(-\frac{r}{\lambda}\bigg)\bigg(1+\frac{r}{\lambda}\bigg)
\cos f\nonumber\\
& & +\frac{2}{na^{2}}\frac{\mu}{r}\alpha\exp\bigg(-\frac{r}{\lambda}\bigg)\bigg(1+\frac{r}{\lambda}\bigg).
\end{eqnarray}
It is worth mentioning that, in the work by \cite{Haranas2011}, the authors derived the same equations but their results are quite different from above ones. Especially, they found the orbital inclination $i$ varies due to the existence of the Yukawa force [Eqs.(18) and (33) in \cite{Haranas2011}]. Given the fact that the Newtonian gravity with the Yukawa correction is a special case of central force so that the angular momentum per unit mass of the 2-body system $\bm{h}$ is conserved as a constant vector \citep{Danby1962,Goldstein2002}, the inclination $i \equiv \arccos (\bm{h}\cdot\hat{\bm{z}}/|\bm{h}|)$, where $\hat{\bm{z}}$ is the unit vector along $z$-axis of the inertial reference system, will never change.

For secular evolution of the orbit, we need to average $R$ over one orbital revolution of the satellite that is
\begin{equation}
  \label{barR}
  \bar{R}  = \frac{1}{T}\int_0^TR \mathrm{d}t =   \frac{\alpha\mu}{a} \exp\bigg(-\frac{a}{\lambda}\bigg) I_0\bigg(\frac{ae}{\lambda}\bigg),
\end{equation}
where $I_0(z)$ is the modified Bessel function of the first kind \citep{Arfken2005}. Therefore, we can have the long term variations of the orbital elements as
\begin{eqnarray}
  \label{}
  \bigg<\frac{\mathrm{d} a}{\mathrm{d} t}\bigg> & = & 0,\\
  \bigg<\frac{\mathrm{d} e}{\mathrm{d} t}\bigg> & = & 0,\\
  \bigg<\frac{\mathrm{d} i}{\mathrm{d} t}\bigg> & = & 0,\\
  \bigg<\frac{\mathrm{d} \Omega}{\mathrm{d} t}\bigg> & = & 0,\\
  \label{seculardomegadt}
  \bigg<\frac{\mathrm{d} \omega}{\mathrm{d} t}\bigg> & = & \alpha \frac{n a\sqrt{1-e^2}}{e\lambda}\exp\bigg(-\frac{a}{\lambda}\bigg) I_1\bigg(\frac{ae}{\lambda}\bigg),\\
  \bigg<\frac{\mathrm{d} M}{\mathrm{d} t}\bigg> & = & n+\alpha(a+\lambda)\frac{2n}{\lambda}\exp\bigg(-\frac{a}{\lambda}\bigg)I_{0}\bigg(\frac{ae}{\lambda}\bigg)\nonumber\\
&&-\alpha\frac{na(e^{2}+1)}{e\lambda}\exp\bigg(-\frac{a}{\lambda}\bigg)I_{1}\bigg(\frac{ae}{\lambda}\bigg),
\end{eqnarray}
where $I_1(z) = \mathrm{d}I_0(z)/\mathrm{d}z$ \citep{Arfken2005}. Our results Eqs. (\ref{barR}) and (\ref{seculardomegadt}) identically match those given by \cite{Iorio2012}. With the dynamics of the satellite under the Yukawa perturbation, we can study its effects on the time transfer of the onboard clock.

\section{Yukawa effects on time transfer}

\label{YukawaTT}

For a satellite surrounding the Earth, the Geocentric Coordinate Time (TCG) is the coordinate time to describe its dynamics and the time variable in its equations of motion \citep{Soffel2003,Kopeikin2011,Soffel2013}. However, the actual reading of a clock onboard the satellite is the proper time $\tau$ \citep{MTW,Soffel2003,Kopeikin2011,Soffel2013}. They diverges with time. To trace the readings of a space-borne clock to a clock on Earth or other time scales, the transformation between TCG and $\tau$ is usually taken as an intermediate.

According to the relativistic theory of time scales \citep{Soffel2003,Kovalevsky2004,Kopeikin2011,Soffel2013}, the transformation between $\tau$ and $t\equiv \mathrm{TCG}$ is
\begin{eqnarray}
\label{dtaudTCG}
\Delta t = \int^{B}_{A}\bigg(1+\frac{1}{c^{2}}V+\frac{1}{2}\frac{v^{2}}{c^{2}}\bigg)d\tau + \mathcal{O}(c^{-4}),
\end{eqnarray}
where, if taking the Yukawa correction into account,
\begin{equation}
  \label{gpV}
  V=\frac{\mu}{r}\bigg[1+\alpha\exp\bigg(-\frac{r}{\lambda}\bigg)\bigg].
\end{equation}
In the above expression of $V$, the contributions from the Earth's oblateness and external bodies are not metioned and we assume they can be properly modelled to adequate precision. (We will go back to this issue in the next section of observability.) Following the procedures used by \cite{Nelson2011}, we can obtain
\begin{eqnarray}
  \label{tauTCG}
  \Delta t & = & \bigg[1+\frac{3}{2}\frac{1}{c^{2}}\frac{\mu}{a}+\alpha\frac{1}{c^{2}}\frac{\mu}{a}\exp\bigg(-\frac{a}{\lambda}\bigg)\bigg]\Delta\tau\nonumber\\
  & & +\frac{2}{c^{2}}\sqrt{\mu a}e\sin E\nonumber\\
  & & +\frac{\alpha}{c^{2}}\sqrt{\mu a}e\exp\bigg(-\frac{a}{\lambda}\bigg)\bigg(1+\frac{a}{\lambda}\bigg)\sin E\nonumber\\
  & & +\mathcal{O}(e^2c^{-2},c^{-4}),
\end{eqnarray}
in which $E$ is the eccentric anomaly. When $\alpha=0$, it returns to the result of \cite{Nelson2011}. From the Eq. (\ref{tauTCG}), we can see that the Yukawa effects on the time transfer have two components: a secular term and a periodic term. The secular term might not be observable because it is mixed with the intrinsic frequency drift of the clock. The periodic term, whose frequency is as same as the leading term due to general relativity, might show up, but its observability depends on its amplitude
\begin{equation}
  \label{AY}
  A_{\mathrm{Y}} = \frac{\alpha}{c^{2}}\sqrt{\mu a}e\exp\bigg(-\frac{a}{\lambda}\bigg)\bigg(1+\frac{a}{\lambda}\bigg).
\end{equation}

\section{Observability of Yukawa effects}

\label{obs}

This section will be dedicated to an important issue: the observability of these Yukawa effects. We focus on a relative small astronomical length scale that the space vehicle stay in the vicinity of Earth and take the domain of Yukawa parameters as 
\begin{equation}
  \label{}
  \mathcal{D}=\{(\lambda,\alpha)|10^6\,\mathrm{m} \le \lambda \le 10^{10}\,\mathrm{m},\,10^{-12}\le\alpha\le10^{-8}\}
\end{equation}
based on the literature [such as \cite{Fischbach1999,Adelberger2003,Deng2009,Lucchesi2010,Deng2011,Lucchesi2011,Iorio2012}]. We adopt a satellite with $a=10^7$\,m and $e=0.01$ as an example.

\begin{figure}
 \centering
  \includegraphics[width=8cm, angle=0]{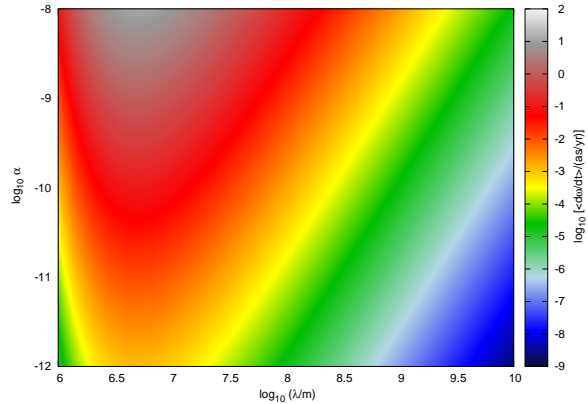}
  \caption{Color indexed secular change of $\omega$ according to Eq. (\ref{seculardomegadt}) in $\mathcal{D}$.}
  \label{Fig:domegadt}
\end{figure}

According to Eq. (\ref{seculardomegadt}), the domain $\mathcal{D}$ can generate the secular change of $\omega$, $\left<\dot{\omega}\right>$, ranging from several nano-arcsecond per year to hundreds arcsecond per year (see color indexed Fig. \ref{Fig:domegadt}). If the $\left<\dot{\omega}\right>$ can be determined to the level of micro-arcsecond per year by laser ranging that the  current best fitted values can reach better than $\sim$0.1 milli-arcsecond per year \citep{Lucchesi2010,Lucchesi2011}, most regions of the Fig. \ref{Fig:domegadt} will be brought into the detectable threshold.

\begin{figure}
 \centering
  \includegraphics[width=8cm, angle=0]{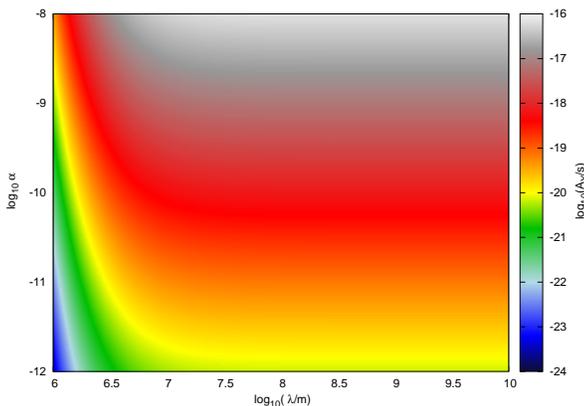}
  \caption{Color indexed amplitude $A_{\mathrm{Y}}$ according to Eq. (\ref{AY}) in $\mathcal{D}$. }
  \label{Fig:AY}
\end{figure}

For the time transfer, the amplitude $A_{\mathrm{Y}}$ [Eq. (\ref{AY})] of the periodic term due to the Yukawa correction holds the value from $\sim 10^{-24}$\,s to $\sim 10^{-16}$\,s in the domain $\mathcal{D}$ (see color indexed Fig. \ref{Fig:AY}). Even if a current optical clock with a fractional frequency inaccuracy of $8.6\times 10^{-18}$ \citep{Chou2010} would be carried onboard the satellite, it is hardly able to detect the Yukawa signal in the duration of one orbit. Since picking up a periodic term usually needs to sample the data in the time interval comparable with its period, which is $\sim 10^4$\,s in this case, such a long time will make the uncertainty of the clock's frequency exceed the Yukawa signal. One of possible ways to improve the situation with the given state of clocks is to dig the higher frequency periodic terms (but perhaps with much less amplitudes) if exist.

At the frequency stability level of a few parts in $10^{18}$, a very important issue involved in the time transfer is the contributions [in Eq. (\ref{gpV})] from the Earth's oblateness and the gravitational tidal fields of external bodies, at least for the Moon and the Sun. They have been well investigated and formulated in \citet{Blanchet2001A&A370.320}, \citet{Soffel2003} and \citet{IERSC2010} and they need to be carefully taken care in order to pick up the Yuakawa signals.

\section{Conclusions and discussion}

\label{condis}

In this work, we investigate the Yukawa effects on the clock onboard a drag-free satellite, including the dynamics of the space vehicle and the time transfer of the clock. We find the Yukawa signal in the time transfer with current state of clocks is much harder to detect than those in the dynamics, especially the secular change of $\omega$ \citep{Lucchesi2010,Lucchesi2011}, by laser ranging in the case of a satellite with $a=10^7$\,m and $e=0.01$.

However, some open issues remain. One of them is whether the Yukawa effect will reach the detectable realm in high frequency terms in time transfer. After all, Eq. (\ref{tauTCG}) is truncated and other terms do exist but perhaps with much less amplitudes. Another interesting issue might be these effects on the interplanetary time transfer link \citep{Deng2012}, which need to step into the region with a larger astronomical scale to length.

\section*{Acknowledgments}

The authors especially would like to thank Professor Tian-Yi Huang of Nanjing University for his fruitful discussions. The work of X.-M.D. is funded by the Natural Science Foundation of China under Grant No. 11103085. The work of Y.X. is supported by the National Natural Science Foundation of China Grant No. 11103010, the Fundamental Research Program of Jiangsu Province of China Grant No. BK2011553, the Research Fund for the Doctoral Program of Higher Education of China Grant No. 20110091120003 and the Fundamental Research Funds for the Central Universities No. 1107020116. This project/publication was made possible through the support of a grant from the John Templeton Foundation. The opinions expressed in this publication are those of the authors and do not necessarily reflect the views of the John Templeton Foundation. The funds from the John Templeton Foundation were provided by a grant to The University of Chicago which also managed the program in conjunction with National Astronomical Observatories, Chinese Academy of Sciences. X.-M. D. appreciates the support from the group of Almanac and Astronomical Reference Systems in the Purple Mountain Observatory of China.

\bibliographystyle{mn2e.bst}
\bibliography{gravityMN20130305.bib}

\end{document}